# Towards femtosecond on-chip electronics

C. Karnetzky[1,2], P. Zimmermann[1,2], C. Trummer[1,2], C. Duque-Sierra[1,2], M. Wörle[3], R. Kienberger[3], and A.W. Holleitner[1,2]

[1] *Walter Schottky Institute and Physics Department, Technical University of Munich, Am Coulombwall 4a, 85748 Garching, Germany.*

[2] *Nanosystems Initiative Munich (NIM), Schellingstr. 4, 80799 Munich, Germany.*

[3] *Physik-Department E11, Technical University of Munich, James-Franck-Str. 1, 85748 Garching. Germany.*

**To combine the advantages of ultrafast femtosecond optics with an on-chip communcation scheme, optical signals with a frequency of several hundreds of THz need to be down-converted to coherent electronic signals of GHz or less. So far, this hasn't been achieved because of the impedance mismatch within electronic circuits and their overall slow response-time. Here, we demonstrate that 14 fs optical pulses in the near-infrared (NIR) can drive electronic on-chip circuits with a bandwidth up to 10 THz. The corresponding electronic pulses propagate in microscopic striplines on a millimeter scale. We exploit femtosecond photoswitches based on tunneling barriers in nanoscale metal junctions to drive the pulses. The non-linear ultrafast response is based on a combination of plasmonically enhanced, multi-photon absorption and quantum tunneling, and gives rise to a field emission of ballistic electrons propagating across the nanoscale junctions. Our results pave the way towards femtosecond electronics integrated in waferscale quantum circuits.**

In 1928 Fowler and Nordheim calculated the electron emission from a metal surface into vacuum based on the theory of Schottky.[1] Very recent work on sub-cycle femtosecond lasers suggests that this electron emission may be exploited for driving ultrafast, coherent currents in nanoscale circuits.[2,3,4] The underlying mechanisms are a photoemission process after multi-photon excitation or a direct tunneling, when the electric field of the laser pulse reduces the surface tunneling barrier.[5,6,7] The electron emission can be significantly increased by utilizing resonant plasmonic nanostructures such as bowties antennas.[3,4,8,9] However, driving a macroscopic unipolar current in such structures requires a symmetry breaking of the spatio-temporal electron dynamics.[10] So far, this has been achieved by using sub-cycle femtosecond pulses for the optical excitation[2,4,11] or by applying strong dc electric fields at the emitter electrodes.[3]

We demonstrate ultrafast, unipolar photoemission currents in asymmetric, plasmonic nanojunctions. The plasmonic junctions are resonantly excited by near-infrared (NIR) femtosecond pulses. The photoemission currents comprise electrons traveling ballistically across the nanojunctions in vaccuum. We demonstrate that this technology can be utilized as an ultrafast photoswitch driving on-chip THz circuits. The ultimate switching time is limited by the laser pulse duration and the time-of-flight of the ballistic electrons, which was reported to be as fast as 900 as for a 8 nm junction gap.[4] This (sub-) femtosecond timescale outperforms known Auston switch technologies based on non-radiative carrier capture sites in semiconductors by two orders of magnitude.[12,13,14,15] Our work reveals that the photoemission dynamics in asymmetric nanojunctions allow to convert a femtosecond NIR-optical pulse into a coherent on-chip signal in the THz range. In this respect, we expand on-chip electronics from GHz up to 10 THz, covering the so-called THz-Gap between electronic and optical applications.[16] The THz pulses are driven by the non-linear photoemission response of the asymmetric nanojunctions, and are coupled into macroscopic co-planar striplines by near-field interactions. The on-chip signal propagation along the striplines extends up to several hundreds of micrometers. In this respect, we show that femtosecond electronics based on asymmetric nanoscale junctions may prove useful for on-chip clock and synchronization dynamics up to 10 THz and to realize a macroscopic on-chip signal transduction on a femtosecond time-scale.

We fabricate asymmetric nanojunctions by focused-ion beam (FIB) milling of a 35 nm thick Au layer on a sapphire substrate with 2 nm Ti as an adhesion layer. Each nanojunction consists of a triangular-shaped emitter ('E') electrode and a plane collector ('C') electrode separated by a vacuum gap with a distance of $d_{gap}$ ~ 90 nm (Fig. 1a). The nanojunctions are positioned in-between two co-planar THz-striplines made from Ti/Au with the emitter and the collector electrodes directly connected to the striplines (Fig. 1b). We use near-infrared (NIR) broadband pulses with a photon energy $E_{pump}$ = (0.9 - 1.3) eV as an excitation of a photocurrent (cf. supplementary Fig. S1). The time-integrated photocurrent $I_{emission}$ is measured between the striplines at zero bias ($V_{bias}$ = 0 V) (Fig. 1c). For time-resolved measurements, an additional NIR probe-pulse at $E_{probe}$ = 1.59 eV and a pulse duration of 100 fs full width at half maximum (FWHM) triggers a semiconductor photoswitch for the electronic read-out with a switching time of about 500 fs.[17,18,19] All measurements are performed at 77 K in vacuum.

Fig. 1b depicts an SEM image of the striplines with the nanojunctions located in the center. The graph is overlaid with a map of $I_{emission}$ of the same area. We find that the maximum current is located at the position of the nanojunctions. The photocurrent $I_{emission}$ is unipolar and its polarity is such that electrons propagate from the emitter to the collector with an amplitude of 360 fA. We note that the spatial extension of $I_{emission}$ is significantly smaller than the laser spot of ~7.5 μm (FWHM) pointing towards a super-linear intensity dependence. Consistently, Fig. 1d shows that $I_{emission}$ follows a power law $(E_{pulse})^\beta$ with a fitted power coefficient in the range of $2 \leq \beta \leq 3$ for the investigated samples. Such coefficients have been reported before[20] and are typically explained by a combination of multi-photon processes and a Fowler-Nordheim tunneling.[7]

We use an SF10-prism compressor to control the tempo-spatial shape of the pump pulse and a second harmonic generation frequency resolved optical gating technique (SHG-FROG) to characterize it.[21] The upper panels of Fig. 1e show the second-harmonic-intensity $\hat{I}_{shg\text{-}frog}$ vs time delay for three different compressor settings. Gaussian-fits yield a FWHM between 19 fs (triangle) and 26 fs (circle), up to ~27 fs for multi-mode pulses (square). The width of shortest pulse (triangle) translates to a temporal FWHM of 14 fs (cf. Supplementary Fig. S1). The lower panel of Fig. 1e shows the corresponding emission current $I_{emission}$ across the asymmetric nanojunctions for the investigated range of compressor positions. We observe a maximum $I_{emission}$ for the shortest laser pulses, while the pulse energy is constant for all compressor positions ($E_{pulse}$ = 150 pJ). In this respect, Fig. 1e demonstrates that $I_{emission}$ depends predominantly on the electric field of the impinging photons instead of the average laser intensity.

For the highest $E_{pulse}$ = 320 pJ and a Gaussian pulse length $\tau_{pulse}$ = 14 fs, we estimate the peak electric field of the pump laser pulse to be $F_{pump}$ = 0.5 Vnm$^{-1}$. For this electric field strength, Fig. 2a shows the corresponding schematic energy diagram of a gold-vacuum interface (black line). As depicted by the colored lines, the potential barrier is reduced by the Schottky effect to

$$W_{barrier} = W_{gold} - \operatorname{Sqrt}[e^3 g F_{pump} / (4\pi\varepsilon_0)], \qquad (1)$$

with $W_{gold}$ = 5.1 eV the work function of gold, $\varepsilon_0$ the vacuum permittivity, $e$ the electron charge, and $g$ the field-enhancement factor. Fig. 2b shows a close-up SEM image of the emitter and collector structure. For $E_{pump}$ = 1.3 eV, we numerically calculate a maximum plasmonic field-enhancement factor $g$ ~4.5 at the tips of the emitter, while at the collector $g < 2$ (cf. inset of Fig. 2b). This asymmetry of the nanojunctions favors the photoemission of electrons from the emitter to the collector, which explains the unipolar amplitude of $I_{emission}$ in our experiments. The amplitude of $I_{emission}$ further translates to ~0.08 electrons emitted per optical pulse in average, which is consistent with a serial emission process as suggested by Fig. 2a.

The photoemission process is described by two distinct mechanisms.[22,23,24] The first covers the absorption of multiple photons with a combined energy of $E_{multi-photon} = \beta \cdot E_{pump} = (1.8 - 3.9)$ eV. This energy range is consistent with the sketched barrier heights in Fig. 2a. For instance, we compute $W_{barrier}$ = 3.2 eV according to Eq. (1) for $g = 5$ at the emitter tips, which nicely explains the measured power coefficient $\beta = 2.1$ in Fig. 1d. A second possible explanation is the Fowler-Nordheim tunneling at $E_{fermi}$ (dashed horizontal arrow in Fig. 2a).[1,7,24] Moreover, a combination of both mechanisms can give rise to multi-photon induced tunneling processes at higher electron energies.[6,7] However, for the highest laser field of $F_{pump}$ = 0.5 Vnm$^{-1}$ and $W_{barrier}$ = 3.2 eV, the so-called Keldysh parameter $\gamma$ can be estimated to be 4.3 in our experiment.[25] This value suggests that the multi-photon absorption is the dominating mechanism in our asymmetric nanojunctions for the given laser pulses.

We design the geometry of the nanojunctions to be resonant to the laser spectrum in order to maximize the photon absorption. Fig. 2c shows the simulated scattering cross section (SCS) vs $E_{pump}$ as well as the experimentally determined extinction of the utilized nanojunctions. We find maxima for both the emitter 'E' and the collector 'C'. As is typical for such nanojunctions, their extinction strongly depends on the polarization of the exciting laser field. We find a maximum extinction for the linear optical polarization aligned along the tips of the emitters in agreement with a dominant dipolar excitation. Consistently, the emission current $I_{emission}$ follows this polarization dependence (cf. Fig 2d).

In the next step, we show that $I_{emission}$ can drive THz-pulses in stripline circuitries as sketched in Fig. 1c. After the excitation of the nanojunctions, the THz-pulses run along the striplines up to several hundreds of micrometers, and are detected on-chip by the time-delayed optical probe pulse in combination with a semiconductor Auston switch.[12,17,18,19] The latter is made from ion-implanted amorphous silicon with a sub-picosecond (~500 fs) time resolution.[17,18,19] The resulting current $I_{transient}$ across the Auston switch is sampled as a function of the time delay $\Delta t$ between the pump and the probe pulse. Importantly, we find a non-linear power dependence of $I_{transient}$ vs $E_{pulse}$ wrt. the pump laser, when the nanojunctions are optically excited (Fig. 3b). The observed power law coefficient $\beta$ coincides with the one deduced for $I_{emission}$ for our junctions (Fig. 1d), which demonstrates that the signal $I_{transient}$ is a global read-out of $I_{emission}$ along the striplines.

The signal $I_{transient}$, as measured at the Auston switch, is directly proportional to the electric field component of the THz-pulse as it propagates along the striplines with macroscopic dimensions (Fig. 3c).[12,13,26] The dispersion of the striplines allows the propagation of signals up to several THz (Fig. 3d), before losses predominantly to the $Al_2O_3$-substrate set in.[27] Assuming an initial Gaussian THz-pulse at the position of the nanojunctions, we accordingly calculate its time- and space-evolution along the striplines.[27] At the position of the semiconducting Auston switch, the computed pulse agrees well with the measured $I_{transient}$ (red line in Fig. 3a). We note that in this calculus, the propagating THz-pulse is convoluted with the (much slower) read-out time of the semiconducting Auston switch. In other words, the time-resolution of our circuit is limited by the charge carrier lifetime of the utilized semiconductor Auston switch.[14] We find an apparent FWHM of the THz-Gaussian to be ~500 fs, which is consistent with the fastest charge carrier lifetimes in ion-implated silicon switches reported so far.[14]

On first view, it is surprising that an optical pulse with 270 THz ($E_{pump}$ ~1.3 eV) can be down-converted to a coherent 2 THz signal in the striplines. However, when electrons propagate ballistically across the nanojunctions, a unipolar displacement current can couple into the THz-striplines by near-field interactions, despite the frequency and momentum mismatch.[12] Fig. 4a depicts side-cuts of the two dominant stripline modes: the odd (even) mode with an opposite (the same) polarity at each stripline (upper vs lower panel). Intriguingly, we can visualize the two modes in our stripline circuits. To do so, we record spatial maps of the maximum $I_{transient}$ for a fixed $\Delta t$ in striplines *without nanojunctions* embedded. Such maps reveal a signal of $I_{transient}$ at all stripline edges at smaller amplitude (Fig. 4b). The polarity distribution of $I_{transient}$ suggests that the odd mode is excited at the center of the striplines (open triangles in Fig. 4b). At the edges of the striplines, the even mode seems to be predominantly excited (filled triangles in Fig. 4b). At all stripline edges, $I_{transient}$ follows an intensity dependence with a power-law close to one or slightly below (Fig. 4c). This suggests that a tunneling process from the striplines into the $Al_2O_3$-substrate dominates, which is reasonable considering the fact that the Ti/Au is directly deposited onto the $Al_2O_3$.[7] We note that in this case, $I_{transient}$ only captures the ultrafast emission processes into the substrate and not the back-flow of the electrons (Supplementary Fig. S2). The latter occurs on timescales of nano- to microseconds and therefore has an amplitude below noise level in our measurements. We point out that the striplines without embedded nanojunctions show no time-averaged $I_{emission}$ because they have a minimum gap distance of 1 µm (open triangles in Fig. 4b) which does not allow charge transfer from one stripline to the other.

In order to maximize the near-field interactions of the nanojunctions and the striplines, we embed the nanojunctions at the center of the striplines and orient the emitter and collector electrodes in a way to favor coupling of the displacement current into the odd mode (cf. Fig. 1a). Fig. 4d shows a spatial map of the maximum $I_{transient}$, and we find a similar spatial distribution as for $I_{emission}$ (cf. Fig. 1b), corroborating the common origin of $I_{transient}$ and $I_{emission}$. In contrast to $I_{emission}$ where we measure the charge current, $I_{transient}$ probes the photoinduced non-equilibrium electric field in the striplines generated by the photoemission processes in the nanojunctions. Given the dispersion up to ~10 THz (cf. Fig. 3d), the time-resolution of $I_{transient}$ is ultimately limited by the ballistic time-of-flight of the electrons from the emitter to the collector.

In conclusion, we demonstrate that asymmetric plasmonic nanojunctions can drive ultrafast photoemission currents at zero bias voltage. These currents drive coherent THz transients in stripline circuits, which can be measured up to a millimeter scale. This presents an encouraging step towards wafer-scale femtosecond electronics. Promising future directions include increasing the fundamental time-resolution by reducing the nanojunctions' gap and the corresponding time-of-flight of the photoemitted electrons and by pushing the stripline circuits to even higher bandwidths, e.g. by shrinking the striplines' dimensions (cf. Fig. 3d for striplines with $w = s = 1$ µm) or by using different substrates including superstrates to further decrease the dispersion and attenuation of the propagating THz-signals.

## Methods

**Fabrication of the asymmetric nanojunctions.** As substrate we use sapphire with a thickness of 430 µm, covered with 300 nm silicon. The silicon is implanted with $O_2$ to yield an excess carrier lifetime of ~500 fs. In a first lithographical step, we etch the silicon using $HF/HNO_3$ to form the Auston switches. In two subsequent optical lithography steps, we first evaporate a Ti/Au film of 2/35 nm for the nanojunctions and then the Ti/Au striplines with 10/300 nm. The asymmetric nanojunctions are fabricated using focused-ion beam (FIB) milling of the Ti/Au film which is located in a distance of ~350 µm to the Auston switch. The striplines have a total length of ~48 mm and are separated by 10 µm.

**On-chip time-domain terahertz spectroscopy.** We use an Er-fiber based pulsed laser (repetition rate 80 MHz) as pump and probe. The pump pulses pass through a non-linear fiber and two SF10-prism pairs to tune the broadband spectrum (0.9 – 1.3) eV as well as the pulse length (>14 fs) with a maximum average laser power of 55 mW. The pump pulses are focused on the nanojunctions by a $CaF_2$ lens ($f$ = 40 mm) to a spot size of ~7.5 µm (FWHM). We achieve identical experimental results also with a refractive objective. The probe pulses have a pulse duration of 100 fs, energy of 1.59 eV, laser power of 80 mW, and are focused on the Auston switch by a x10 objective. The spot size of the probe laser is chosen in a way to maximize the read out signal and yields 4 µm. After excitation by the pump pulse, the ultrafast field emission current across the asymmetric nanojunctions couple into the striplines. Consequently, an electromagnetic transient, proportional to the initial current, propagates along the striplines. After a time delay $\Delta t$, the probe pulse triggers the Auston switch. The presence of the electromagnetic transient at the switch drives the current $I_{transient}(\Delta t)$ to the read out contact. We use frequency modulation of the pump laser in combination with a lock-in amplifier for the read out. All measurements are done in vacuum ($10^{-6}$ mbar) at $T$ = 77 K.

**Simulation of plasmonic enhancement of the nanojunctions.** To simulate the field enhancement of the nanojunctions, we apply finite element simulations using COMSOL Multiphysics®. The model for the simulation in Fig. 2b consists of a gold nanojunction with height 37 nm on a sapphire substrate. We calculate the scattering cross section of an incident light as well as the generated electric field distribution in the nanojunctions to deduce the field enhancement at the emitter tips and the collector.

**Simulation of the THz dispersion.** To calculate the THz dispersion, we use COMSOL 3D frequency domain and time domain simulations of striplines with height $h$ = 300 nm and different values for the width $w$ and separation $s$ (cf. Fig. 3d). We simulate the propagation of plane waves with different frequencies along the striplines to get the dispersion relation of the effective refractive index $n_{eff}$ and the attenuation.

**Acknowledgements**

We acknowledge S. Govorov and A. Bandrauk for very helpful discussions and J. Klein for technical support. The work was supported by the European Research Council (ERC) under Grant NanoREAL (No. 306754).

**Author contributions**

C.K. and A.W.H. designed the experiments, C.K., P.Z., C.T., C.D, M.W., R.K., A.W.H performed the experiments, analyzed the data, and wrote the manuscript. All authors have given approval to the final version of the manuscript.

**Competing financial interests**

The authors declare no competing interests.

**Supporting Information Available**

Supplementary Information accompanies this paper at www.nature.com. Reprints and permission information is available online at http://npg.nature.com/reprintsandpermissions/.

Correspondence and requests for materials should be addressed to A.W.H.

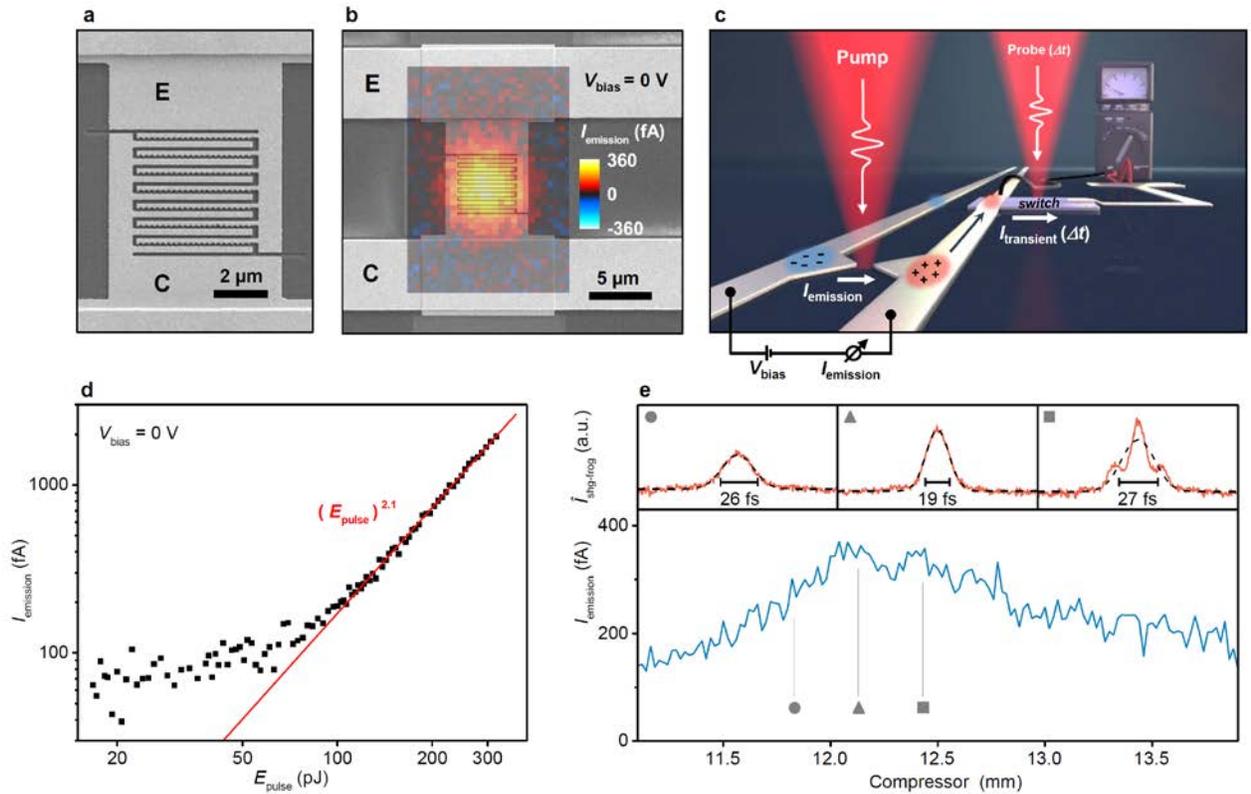

**Figure 1 | Femtosecond photoemission in nanoscale junctions and THz on-chip circuits.**
**a,** Scanning electron microscope (SEM) image of Ti/Au-contacts with asymmetric nanojunctions with the emitter (collector) denoted as 'E' ('C'). **b,** Lateral map of the unipolar photoemission current $I_{emission}$ at zero bias $V_{sd}$ across asymmetric nanojunctions (inner graph), which are contacted by two Ti/Au-striplines (outer SEM image). **c,** On-chip THz time-domain circuit with optical femtosecond pump- and probe-pulses triggering the electronic read-out. $I_{emission}$ describes the time-integrated current, while $I_{transient}$ captures the time-resolved electromagnetic transients in the striplines at a time-delay $\Delta t$. **d,** Non-linear $I_{emission}$ vs laser pulse energy $E_{pulse}$ with a power law fit (red line). **e,** Lower graph shows $I_{emission}$ vs laser compressor position at a fixed $E_{pulse}$ = 150 pJ. The three upper insets show the second harmonic generation frequency resolved optical gating (SHG-FROG) intensity for three given compressor positions denoted by a circle, triangle, and square. All measurements are performed at 77 K.

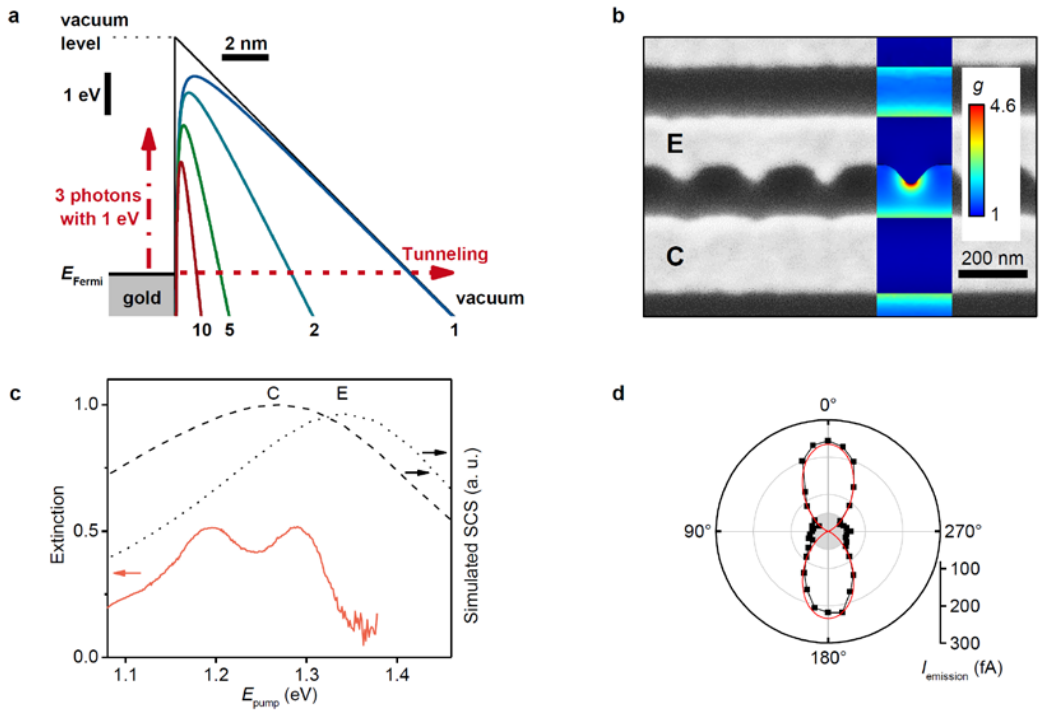

**Figure 2 | Asymmetric nanoscale junctions for plasmonically enhanced photoemission.**
**a,** Schematic energy diagram of the gold-vacuum interface at the emitter at an electric field of $F_{pump} = 0.5$ Vnm$^{-1}$ (black line). The Fermi energy $E_{Fermi}$ is ~5.1 eV below the vacuum level (dotted line). The barrier can be overcome by a multi-photon absorption (dashed dotted line) or a tunneling process (dashed line). The colored lines consider the Schottky effect and a field-enhancement of 1 (blue), 2 (turquois), 5 (green), and 10 (red). **b,** SEM image of asymmetric nanojunctions with emitter ('E') and collector ('C') electrodes. Inset: numerically computed field enhancement $g$ within such an asymmetric nanojunction for $E_{pump} = 1.3$ eV. At the emitter tips, we compute a maximum field-enhancement factor $g \sim 4.6$. **c,** Simulated scattering cross section (SCS) of the emitter (collector) 'E' ('C') vs $E_{pump}$ compared to the normalized, experimentally determined extinction of such nanojunctions. **d,** Polarization-dependence of $I_{emission}$ at $E_{pulse} = 150$ pJ and a FWHM of 14 fs of the pump pulse. Red line is a cosine fit, the gray area indicates the noise level.

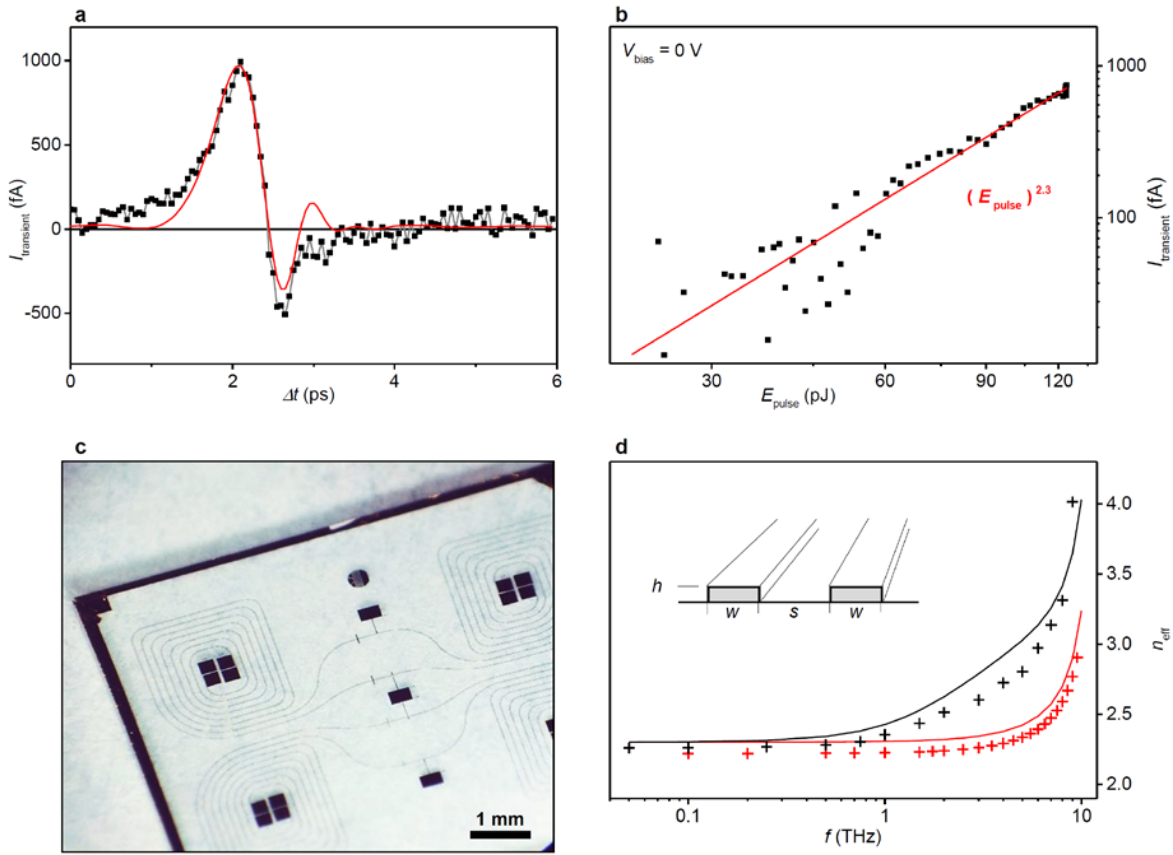

**Figure 3 | Non-linear THz-pulses in macroscopic on-chip circuits**. **a,** Time-resolved $I_{transient}$ vs $\Delta t$ (black) and fit function (red) after exciting a nanojunction integrated in the stripline circuits with a 14 fs laser pulse at $E_{pulse}$ = 124 pJ. The THz-signal is detected after a propagation length of 300 µm. **b,** Non-linear $I_{transient}$ vs $E_{pulse}$ with a power law fit (red line) showing a similar power law coefficient $\beta$ as found for $I_{emission}$ (cf. Fig. 1d). **c,** Microscope image of the utilized THz-circuitries on a sapphire chip. **d,** Dispersion relation of the effective diffraction index $n_{eff}$ of coplanar gold striplines on a sapphire substrate: black (red) crosses depict numerical simulations with dimensions $h$ = 300 nm, $w$ = 5 µm (1 µm), and $s$ = 10 µm (1 µm). The black (red) line shows analytical solutions for symmetrically spaced striplines with dimensions: $h$ = 300 nm, $w = s$ = 10 µm (1 µm).[28]

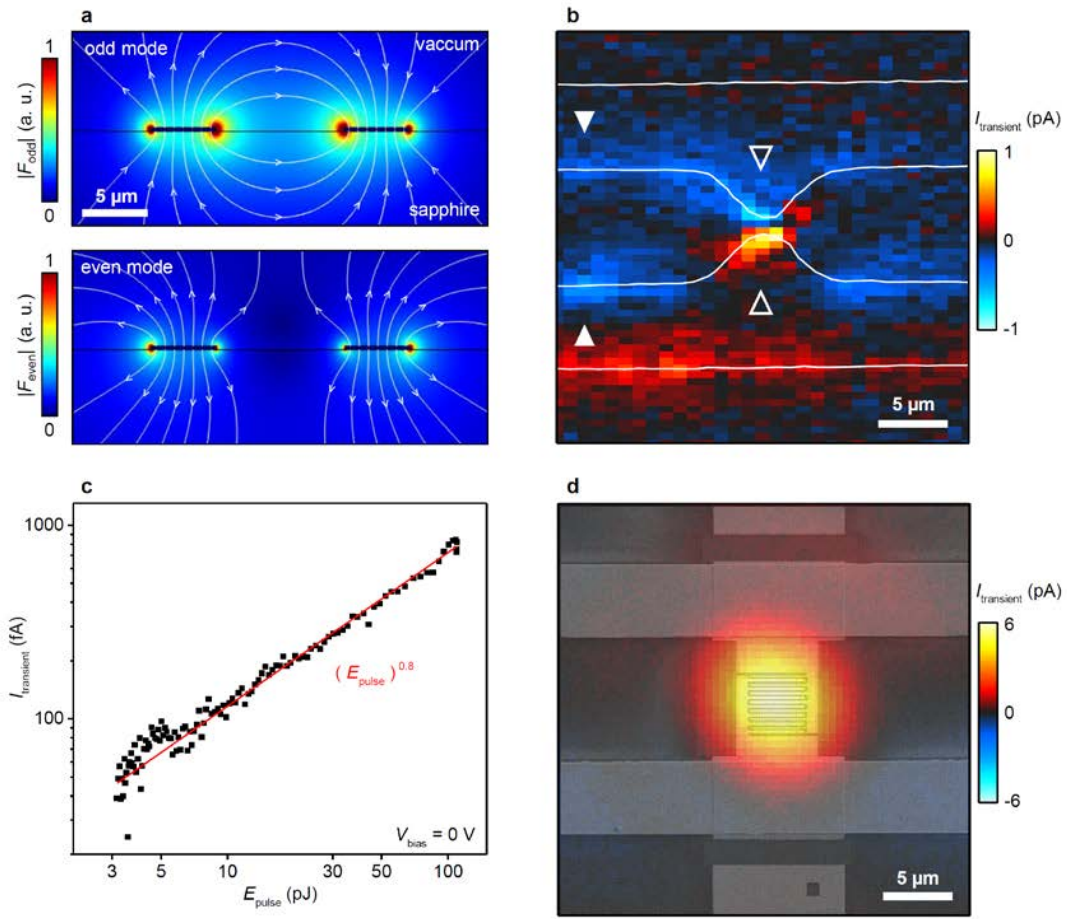

**Figure 4 | Femtosecond near-field coupling of NIR pulses to THz stripline modes. a,** Simulated electric field distribution of the coplanar striplines (black) with the odd (even) mode in the upper (lower) panel. Color code describes the absolute electric field. The arrows denote the direction of the electric field vector in the image plane. **b,** Spatial map of $I_{transient}$ at fixed $\Delta t$ for a sample without nanojunctions (white lines indicate the striplines). The odd mode is excited in the center with a minimum distance of 1 µm between the striplines (open triangles). The even mode is excited at edges where the striplines have a distance of 10 µm (filled triangles). **c,** $I_{transient}$ vs. $E_{pulse}$ for striplines without nanojunctions showing an almost linear dependence (red line). **d,** Spatial map of $I_{transient}$ for asymmetric nanojunctions integrated in the striplines with an overlaid SEM image.

# Towards femtosecond on-chip electronics

C. Karnetzky[1,2], P. Zimmermann[1,2], C. Trummer[1,2], C. Duque-Sierra[1,2], M. Wörle[3], R. Kienberger[3], and A.W. Holleitner[1,2]

[1] *Walter Schottky Institute and Physics Department, Technical University of Munich, Am Coulombwall 4a, 85748 Garching, Germany.*

[2] *Nanosystems Initiative Munich (NIM), Schellingstr. 4, 80799 Munich, Germany.*

[3] *Physik-Department E11, Technical University of Munich, James-Franck-Str. 1, 85748 Garching, Germany.*

- supporting materials –

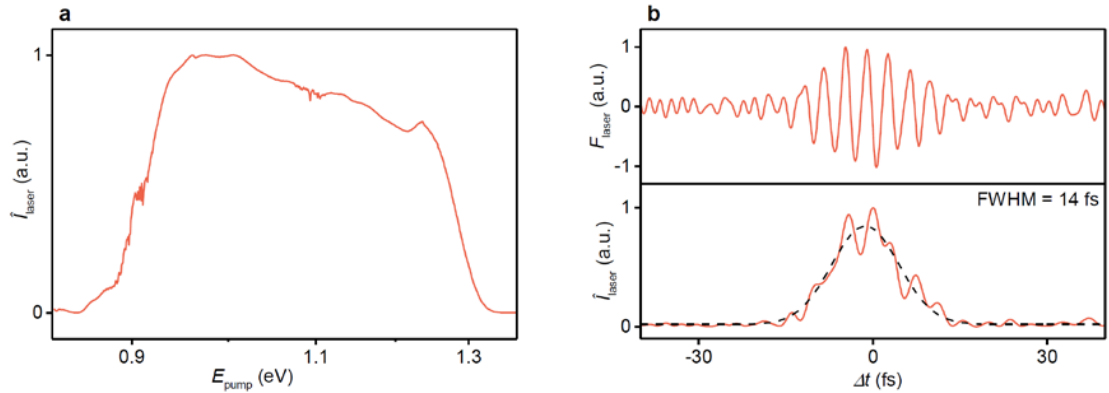

**Supplementary Figure S1 | Frequency resolved optical gating of the utilized laser pulse.**
**a,** Laser intensity $\hat{I}_{laser}$ vs. $E_{pump}$ of the utilized pump laser. **b,** Upper panel (lower panel) shows the electric field $F_{laser}$ (laser intensity $\hat{I}_{laser}$) of the shortest pulses vs time retrieved from the phase-resolved SHG-FROG characterization.

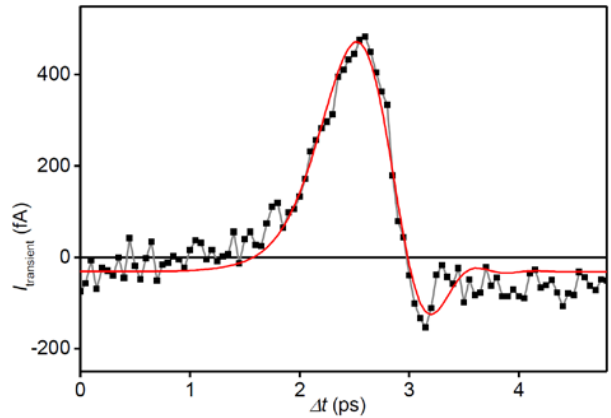

**Supplementary Figure S2 | THz-transient measured at the stripline edges.** $I_{transient}$ vs time delay $\Delta t$ for an excitation position of the pump laser at the edges of the striplines (open triangles in Fig. 4b of the main manuscript). The transient has an initial FWHM of 500 fs, which is consistent with the time-scale given by the non-radiative lifetime of the semiconductor Auston switch.